\documentclass[aps, twocolumn, letterpaper, superscriptaddress]{revtex4}

\usepackage{amsmath}
\usepackage{amssymb}
\usepackage{mathrsfs}
\usepackage[dvipdfmx]{graphicx}
\usepackage{color}
\usepackage{xspace}
\usepackage{bbold}
\usepackage{MnSymbol}

\newcommand{\ie}{{i.e.,\/}\xspace}
\newcommand{\etal}{{\it et~al.\/}\xspace}

\newcommand{\myRef}[1]{Ref.~\cite{#1}}
\ifx\Ref\undefined
\newcommand{\Ref}[1]{\myRef{#1}}
\else
\renewcommand{\Ref}[1]{\myRef{#1}}
\fi

\newcommand{\eq}[1]{(\ref{#1})}
\newcommand{\Eq}[1]{Eq.~(\ref{#1})}

\newcommand{\Fig}[1]{Fig.~\ref{#1}}
\newcommand{\Figs}[1]{Figs.~\ref{#1}}
\newcommand{\Sec}[1]{Sec.~\ref{#1}}
\newcommand{\Refs}[1]{Refs.~\cite{#1}}

\newcommand{\mc}[1]{\mathcal{#1}}
\newcommand{\mcc}[1]{\mathfrak{#1}}
\newcommand{\msf}[1]{\mathsf{#1}}

\newcommand{\pd}{\partial}
\newcommand{\del}{\nabla}
\renewcommand{\vec}[1]{\boldsymbol{\rm #1}}
\newcommand{\oper}[1]{\smash{\widehat{#1}}}
\newcommand{\dd}{\mathrm{d}}

\newcommand{\tored}[1]{{\color{black}#1}}

\newcommand{\Deff}{\msf{D}}
\newcommand{\sigmap}{{\bar{\sigma}}}
\newcommand{\dk}{\pi}

\newcommand{\parade}{{PARADE}\xspace}
\begin{document}

\title{Quasioptical modeling of wave beams with and without mode conversion:\\IV. Numerical simulations of waves in dissipative media}

\author{K. Yanagihara}
\affiliation{Naka Fusion Institute, National Institutes for Quantum and Radiological Science and Technology, 311-0193, Naka, Ibaraki, Japan}

\author{I. Y. Dodin}
\affiliation{Princeton Plasma Physics Laboratory, Princeton, New Jersey 08543, USA}
\affiliation{Department of Astrophysical Sciences, Princeton University, Princeton, New Jersey 08544, USA}

\author{S. Kubo}
\affiliation{National Institute for Fusion Science, National Institutes of Natural Sciences, 509-5292, Toki, Gifu, Japan}

\date{\today}

\begin{abstract}
We report the first quasioptical simulations of wave beams in a hot plasma using the quasioptical code \parade (PAraxial RAy DEscription) [Phys. Plasmas {\bf 26}, 072112 (2019)]. This code is unique in that it accounts for inhomogeneity of the dissipation-rate across the beam and mode conversion simultaneously. We show that the dissipation-rate inhomogeneity shifts beams relative to their trajectories in cold plasma and that the two electromagnetic modes are coupled via this process, an effect that was ignored in the past. We also propose a simplified approach to accounting for the dissipation-rate inhomogeneity. This approach is computationally inexpensive and simplifies analysis of actual experiments.
\end{abstract}

\maketitle
\bibliographystyle{full}

%%%%%%%%%%%%%%%%%%%%%%%%%%%%%%%%%%%%%%%
\section{Introduction}
\label{sec:intro}

Modeling of radiofrequency waves in fusion plasmas requires accurate calculations of their power deposition through various resonant mechanisms. For electron-cyclotron (EC) \cite{ref:bornatici83, ref:erckmann94} and lower-hybrid \cite{ref:bonoli84} waves, which have short enough wavelengths, this is commonly done using geometrical-optics ray tracing \cite{ref:bernstein75, ref:friedland80}. However, this method ignores a number of important effects, including diffraction, so the deposition profiles are often predicted to be more peaked than they are in reality \cite{ref:poli01a}.
To deal with this problem, a number of quasioptical models have been proposed \cite{ref:poli01a, ref:poli01b, ref:poli18, ref:pereverzev98, ref:mazzucato89, ref:nowak93, ref:peeters96, ref:farina07, ref:balakin07a, ref:balakin07b, ref:balakin08a, ref:balakin08b}. However, most of these models \cite{ref:poli01a, ref:poli01b,ref:poli18, ref:pereverzev98, ref:mazzucato89, ref:nowak93, ref:peeters96, ref:farina07} assume that wave beams maintain a particular (Gaussian) transverse profile, and they also ignore variations of the dissipation-rate within the beam cross section.
One exception to this is the model by Balakin \etal \cite{ref:balakin07a, ref:balakin07b, ref:balakin08a, ref:balakin08b}; still, this model assumes single-mode beams and thus cannot describe mode conversion, which is often important \cite{ref:dodin17, ref:tsujimura15, ref:kubo15}. To fill the gap, a more comprehensive quasioptical theory and the corresponding code \parade (PAraxial RAy DEscription) have been developed recently \cite{ref:pop1, ref:pop2, ref:pop3}, and its preliminary applications have already been reported \cite{ref:itc27}.

Here, we report the first applications of \parade to modeling quasioptical wave beams with mode conversion in hot plasma. Our findings indicate that conventional simulations overlook important effects connected with: (i)~the inhomogeneity of the dissipation-rate within the beam cross section and (ii)~the O--X mode conversion. The simulations reported here are the first ones that account for these effects simultaneously. We find that the dissipation-rate inhomogeneity shifts wave beams relative to their trajectories in cold plasma. \parade also predicts that the two plasma modes are coupled via this process, an effect that was ignored in the past. We also propose a simplified approach to accounting for the dissipation-rate inhomogeneity to speed-up the calculations in a practical fusion plasma geometry.

Our paper is organized as follows. In \Sec{sec:theory}, we briefly overview the theoretical model underlying \parade. In \Sec{sec:dissip}, we introduce the two dissipation models that we use, and we also report the coupling between the~X and~O modes caused by the dissipation. In \Sec{sec:conc}, we summarize our main results.

%%%%%%%%%%%%%%%%%%%%%%%%%%%%%%%%%%%%%%%
\section{Theoretical model}
\label{sec:theory}

%--------------------------------------
\subsection{Basic equations}
\label{sec:basic}

To describe the theory that underlies \parade, let us start with the equation for the electric field~$\vec{E}$ of a linear wave governed by a general dispersion operator~$\oper{\vec{D}}$:
\begin{gather}\label{eq:E}
\oper{\vec{D}} \vec{E} = 0.
\end{gather}
We assume that the field is stationary, with constant frequency $\omega$, and has an eikonal form $\vec{E} =  e^{-i \omega t + i\theta(\vec{x})}\vec{\psi}(\vec{x})$. (The time dependence is henceforth omitted for brevity.) Here, the scalar function $\theta$ is a rapidly varying ``reference phase'', $\vec{k} \doteq \del \theta$ is the local wave vector (the symbol $\doteq$ denotes definitions), and the complex vector $\vec{\psi}$ is a slowly varying envelope. We also introduce the following small parameters:
\begin{gather}\label{eq:eps}
\epsilon_\parallel \doteq \lambda/L_\parallel,
\quad
\epsilon_\perp \doteq \lambda/L_\perp,
\quad
\epsilon_\parallel \sim \epsilon_\perp^2 \ll 1,
\end{gather}
where $\lambda \doteq 2\pi/k$ is the wavelength, $L_\parallel$ is the characteristic scale of the beam field along the group velocity at the beam center, and $L_\perp$ is the minimum scale of the field in the plane transverse to the group velocity. The medium-inhomogeneity scale is assumed to be of the same order as $L_\parallel$ or larger. Under these assumptions, \Eq{eq:E} can be expressed as
\begin{gather}\label{eq:psi}
\Deff \vec{\psi} + \oper{\vec{\mc{L}}} \vec{\psi} = 0,
\end{gather}
where the operator $\oper{\vec{\mc{L}}} = \mc{O}(\epsilon_\perp)$ is specified in \Ref{ref:pop1} (also see below), and the matrix $\Deff$ is found from the Weyl symbol of $\oper{\vec{D}}$, or the local dispersion matrix $\vec{D}$ \cite{ref:pop1}, that satisfies the ordering
\begin{gather}\label{eq:DHDA}
\vec{D}_H = \mc{O}(1), \quad \vec{D}_A \leq \mc{O}(\epsilon_\perp).
\end{gather}
The indices $H$ and $A$ denote the Hermitian part and the anti-Hermitian part, respectively. For our purposes, it is sufficient to adopt \cite{ref:pop1}
\begin{gather}
\vec{D}_H(\vec{x}, \vec{p})
= \frac{c^2}{16\pi \omega^2}\,[\vec{p}\vec{p} - (\vec{p} \cdot \vec{p})\mathbb{1}]
+ \frac{1}{16\pi}\,\vec{\varepsilon}_H(\vec{x}, \vec{p}),\\
\vec{D}_A(\vec{x}, \vec{p})
= \frac{1}{16\pi}\,\vec{\varepsilon}_A(\vec{x}, \vec{p}),
\label{eq:DA}
\end{gather}
where $\mathbb{1}$ is a unit matrix and $\vec{\varepsilon}$ is the dielectric tensor found, for example, in \Ref{book:stix}; its dependence on $\omega$ is assumed but not emphasized, since $\omega$ is constant. (Here, $\vec{p}$ denotes any given wave vector, as opposed to $\vec{k}$, which is the specific wave vector determined by $\theta$; see above.)

%--------------------------------------
\subsection{Polarization vectors and matrices}
\label{sec:pol}

Since $\vec{D}_H$ is assumed as the dominant part of the dispersion operator in \Eq{eq:psi}, it is convenient to decompose the envelope $\vec{\psi}$ in the basis $\{\vec{\eta}_s\}$ of the orthogonal eigenvectors of $\vec{D}_H$, \ie $\vec{D}_H \vec{\eta}_s = \Lambda_s \vec{\eta}_s$. This decomposition can be written as follows:
\begin{gather}\label{eq:psia}
\vec{\psi} = \vec{\eta}_{\rm o} a^{\rm o} + \vec{\eta}_{\rm x} a^{\rm x} +
\bar{\vec{\eta}}\bar{a},
\end{gather}
where $a^{\rm o}$, $a^{\rm x}$, and $\bar{a}$ are complex coefficients, $\vec{\eta}_{\rm o}$ and $\vec{\eta}_{\rm x}$ are the polarization vectors of the O- and X-mode in homogeneous plasma, and $\bar{\vec{\eta}}$ is the third eigenvector of $\vec{D}$ that is orthogonal to both of them.

In general, the O and X modes are coupled, which means that both $\Lambda_{\rm o}$ and $\Lambda_{\rm x}$ are close to zero simultaneously and
\begin{gather}
a^{\rm o} = \mc{O}(1), \quad a^{\rm x} = \mc{O}(1), \quad \bar{a} = \mc{O}(\epsilon_\perp).
\end{gather}
The small amplitude $\bar{a}$ can be calculated perturbatively and does not enter quasioptical equations explicitly. Instead, we work with a two-dimensional amplitude vector
\begin{gather}
\vec{a} = \left(
\begin{array}{c}
a^{\rm o}\\
a^{\rm x}
\end{array}
\right)
\end{gather}
and the $3 \times 2$ ``polarization matrix'' $\vec{\Xi}$ that contains the vectors $\vec{\eta}_{\rm o}$ and $\vec{\eta}_{\rm x}$ as its columns,
\begin{gather}\label{eq:Xi}
\vec{\Xi} = \left(
\begin{array}{cc}
\vec{\eta}_{\rm o} & \vec{\eta}_{\rm x}
\end{array}
\right).
\end{gather}
Then, $\vec{\psi}$ can be expressed as follows:
\begin{gather}\label{eq:psiXi}
\vec{\psi} = \vec{\Xi} \vec{a} + \mc{O}(\epsilon_\perp).
\end{gather}

Since we consider the beam dynamics in coordinates that are close to Euclidean, the dual-basis vectors can be adopted in the form $\vec{\eta}^{\rm o} \approx \vec{\eta}_{\rm o}$ and $\vec{\eta}^{\rm x} \approx \vec{\eta}_{\rm x}$, and we also introduce a $2 \times 3$ matrix
\begin{gather}\label{eq:Xidual}
\vec{\Xi}^+ = \left(
\begin{array}{c}
\vec{\eta}^{{\rm o}*} \\ \vec{\eta}^{{\rm x}*}
\end{array}
\right).
\end{gather}
[For more general definitions, see \Ref{ref:pop1}.] As seen easily, this matrix satisfies $\vec{\Xi}^+ \vec{\Xi} = \mathbb{1}$, and
\begin{gather}\label{eq:Lambda}
\vec{\Lambda} \doteq \vec{\Xi}^+ \vec{D}_H \vec{\Xi} = \left(
\begin{array}{cc}
\Lambda_{\rm o} & 0\\
0 & \Lambda_{\rm x}
\end{array}
\right).
\end{gather}

In the single-mode case, also considered in \Refs{ref:balakin07a, ref:balakin07b, ref:balakin08a, ref:balakin08b}, the above equations are simplified. For example, assume that a wave consists mainly of the O mode. (The X-mode case is treated similarly.) Then,
\begin{gather}
a^{\rm o} = \mc{O}(1), \quad a^{\rm x} = \mc{O}(\epsilon_\perp), \quad \bar{a} = \mc{O}(\epsilon_\perp),
\end{gather}
and the polarization matrix becomes $3 \times 1$ dimensional, $\vec{\Xi} = \vec{\eta}_{\rm o}$, so it is just the O-mode polarization vector. Accordingly,
\begin{gather}
\vec{\psi} = \vec{\Xi} a^{\rm o} + \mc{O}(\epsilon_\perp),
\end{gather}
where the correction $\mc{O}(\epsilon_\perp)$ can be found perturbatively but if needed but otherwise is inessential. Similarly, $\vec{\Xi}^+$ is a row vector in this case, namely, $\vec{\Xi}^+ = \vec{\eta}^{\rm o}$. Accordingly, $\vec{\Xi}^+ \vec{\Xi} = 1$ and $\Lambda_{\rm o} \doteq \vec{\Xi}^+ \vec{D}_H \vec{\Xi}$ is a scalar. This single-mode model is used in simulations reported below in Secs.~\ref{sec:full} and \ref{sec:apprx}.

%--------------------------------------
\subsection{Reference ray and new coordinates}
\label{sec:rr}

%Using the relations in \Sec{sec:pol}, \Eqs{eq:psi} and (\ref{eq:DHDA}) yield,
%%
%\begin{gather}\label{eq:Laeta}
%\vec{D}_H \vec{\psi} = \Lambda_s a^s \vec{\eta}_s \approx 0.
%\end{gather}
%%
%When $\Lambda_s$ is small, the corresponding unit eigenvector $\vec{\eta}_s$ can approximately satisfy \Eq{eq:Laeta}, and corresponding amplitudes $a^s$ can have meaning volume.

The evolution of the wave amplitude is considered relative to the a ``reference ray'' (RR) that is governed by Hamilton's equations
\begin{gather}\label{eq:ray}
\frac{\dd \vec{X}}{\dd \zeta} = \frac{1}{V_\star}\,\frac{\pd H_\star}{\pd \vec{K}},
\quad
\frac{\dd \vec{K}}{\dd \zeta} = - \frac{1}{V_\star}\,\frac{\pd H_\star}{\pd \vec{X}},
\end{gather}
where $\vec{X}$ and $\vec{K}$ are the ray coordinate and the ray wavevector, $\zeta$ is the path along the ray trajectory, and $V_\star \doteq |\partial H_\star/\partial \vec{K}|$ is the absolute value of the group velocity. The ray Hamiltonian $H_\star$ is
\begin{gather}\label{eq:Hstar}
H_\star \doteq \frac{1}{2}\,(\Lambda_{\star {\rm o}} + \Lambda_{\star {\rm x}})
\end{gather}
for a mode-converting beam and $H_\star \doteq \Lambda_{\star s}$ for a single-mode beam. Here and further, the index $\star$ denotes that the corresponding quantity is evaluated on the RR.

Next, we introduce the RR-based curvilinear coordinates $\tilde{x}^\mu \equiv \{\zeta, \tilde{\varrho}^1, \tilde{\varrho}^2\}$, where $\tilde{\varrho}^\sigma$ are, loosely speaking, the orthogonal coordinates on the plane transverse to the group velocity of the RR as specified in \cite{ref:pop2}. (Here and further, the indices $\sigma$ and $\sigmap$ span from 1 to 2; other Greek indices span from 1 to 3.) The basis vectors $\tilde{\vec{e}}_\mu$ of the new coordinates ($\dd \vec{x} = \tilde{\vec{e}}_\mu \dd\tilde{x}^\mu$) are defined such that
\begin{gather}\label{eq:eee}
\tilde{\vec{e}}_{\star\mu} \cdot \tilde{\vec{e}}_{\star\nu} = \delta_{\mu\nu},
\quad
\left[\pd \tilde{\vec{e}}_{\sigma}(\tilde{x})/\pd \tilde{\varrho}^\sigmap\right]_\star = 0.
\end{gather}
Then,
\begin{gather}
\vec{x} \approx \vec{X}(\zeta) + \left(
\begin{array}{cc}
\tilde{\vec{e}}_{\star 1} & \tilde{\vec{e}}_{\star 2}
\end{array}
\right)
\left(
\begin{array}{c}
\tilde{\varrho}^1\\
\tilde{\varrho}^2
\end{array}
\right).
\end{gather}

%--------------------------------------
\subsection{Quasioptical equation}
\label{sec:qo}

To simplify the field equation, we introduce the rescaled complex vector amplitude $\vec{\phi} = \sqrt{V_\star} \vec{a}$. Then, as shown in \Ref{ref:pop3}, \Eq{eq:psi} leads to the following parabolic equation:
\begin{align}\label{eq:parab}
\frac{\pd \vec{\phi}}{\pd \zeta} = \frac{1}{V_\star} \biggl[ &
-i( \tilde{\vec{\mcc{L}}}_{\star\sigma\sigmap} \tilde{\varrho}^{\sigma} \tilde{\varrho}^{\sigmap}
  + \tilde{\vec{\mcc{M}}}_{\star\sigma} \tilde{\varrho}^\sigma
  + \vec{M}_\star
  - \vec{U}_\star) \vec{\phi}
\notag\\&
+ \frac{i}{2}\,\tilde{\vec{\Phi}}^{\sigma\sigmap}_\star \pd^2_{\sigma\sigmap} \vec{\phi}
+ \vec{\Gamma} \vec{\phi}
\notag\\&
- ( \tilde{\vec{u}}^\sigma_\star
  + \tilde{\vec{\vartheta}}_\star^\sigma{}_{\sigmap} \tilde{\varrho}^{\sigmap}) \pd_\sigma \vec{\phi}
- \frac{\tilde{\vec{\vartheta}}_\star^\sigma{}_\sigma}{2} \vec{\phi}
\biggr],
\end{align}
In the single-mode case, when $\vec{\phi}$ is a scalar, one similarly has \cite{ref:pop2}
\begin{align}
\frac{\pd \phi}{\pd \zeta} = \frac{1}{V_\star} \biggl[ &
-i(\tilde{\mcc{L}}_{\star\sigma\sigmap} \tilde{\varrho}^{\sigma} \tilde{\varrho}^{\sigmap}
- U_\star) \phi
+ \frac{i}{2}\,\tilde{\Phi}^{\sigma\sigmap}_\star \pd^2_{\sigma\sigmap} \phi
\notag\\&
+ \Gamma \phi
- \tilde{\vartheta}_\star^\sigma{}_{\sigmap} \tilde{\varrho}^{\sigmap} \pd_\sigma \phi
- \frac{\tilde{\vartheta}_\star^\sigma{}_\sigma}{2} \phi
\biggr],
\end{align}
which is an alternative representation of the field equation assumed in \Refs{ref:balakin07a, ref:balakin07b, ref:balakin08a, ref:balakin08b}. Here, $\pd_\sigma \doteq \pd/\pd \tilde{\rho^\sigma}$, summation over repeating indices is assumed, and the coefficients are expressed through $\vec{D}$ as described in \Ref{ref:pop3}. Models for $\vec{\Gamma}$ are discussed in \Sec{sec:dissip} in detail.

%%%%%%%%%%%%%%%%%%%%%%%%%%%%%%%%%%%%%%%
\section{Dissipation in hot plasma}
\label{sec:dissip}

The matrix $\vec{\Gamma}$, which represents dissipation, is given by~\cite{ref:pop3}
\begin{gather}\label{eq:Gamma}
\vec{\Gamma} = \vec{\Xi}_\star^+ \vec{D}_A(\vec{x},\vec{k}(\vec{x})) \vec{\Xi}_\star,
\end{gather}
where $\vec{D}_A$ is given by \Eq{eq:DA} and $\vec{\Xi}_\star$ is the polarization matrix [\Eq{eq:Xi}] evaluated on the RR. (In the single-mode case, $\vec{\Xi}_\star$ becomes a vector, and then $\Gamma$ is a scalar.) In an inhomogeneous medium, $\vec{\Gamma}$ is inhomogeneous, which results in variation of the dissipation-rate within the beam cross section [\Fig{fig:schematic}(a)]. Most quasioptical codes \cite{ref:poli01a, ref:poli01b, ref:poli18, ref:pereverzev98, ref:mazzucato89, ref:nowak93, ref:peeters96, ref:farina07} ignore this fact and adopt
\begin{gather}\label{eq:GammaStar}
\vec{\Gamma} \approx \vec{\Gamma}_\star = \vec{\Xi}_\star^+ \vec{D}_{A\star} \vec{\Xi}_\star
\end{gather}
instead [\Fig{fig:schematic}(b)]. This leads to incorrect predictions for actual heating rates, as discussed in \Ref{note:nf1}. In \parade, we adopt two different models to calculate $\vec{\Gamma}$ more accurately, as discussed in \Sec{sec:full} and \Sec{sec:apprx}. Also note that unlike in conventional single-mode models that treat the dissipation coefficient as a scalar \cite{ref:poli01a, ref:poli01b, ref:poli18, ref:pereverzev98, ref:mazzucato89, ref:nowak93, ref:peeters96, ref:farina07, ref:balakin07a, ref:balakin07b, ref:balakin08a, ref:balakin08b, ref:pop2}, our $\vec{\Gamma}$ [\Eq{eq:Gamma}] is generally nondiagonal, so it couples $\phi^{\rm o}$ and $\phi^{\rm x}$ in \Eq{eq:parab}. This effect, which we call dissipation-driven mode conversion, is discussed in \Sec{sec:mc}.

%--------------------------------------
\subsection{Exact dissipation matrix}
\label{sec:full}

\begin{figure*}
\begin{center}
\includegraphics[width=15.0cm,clip]{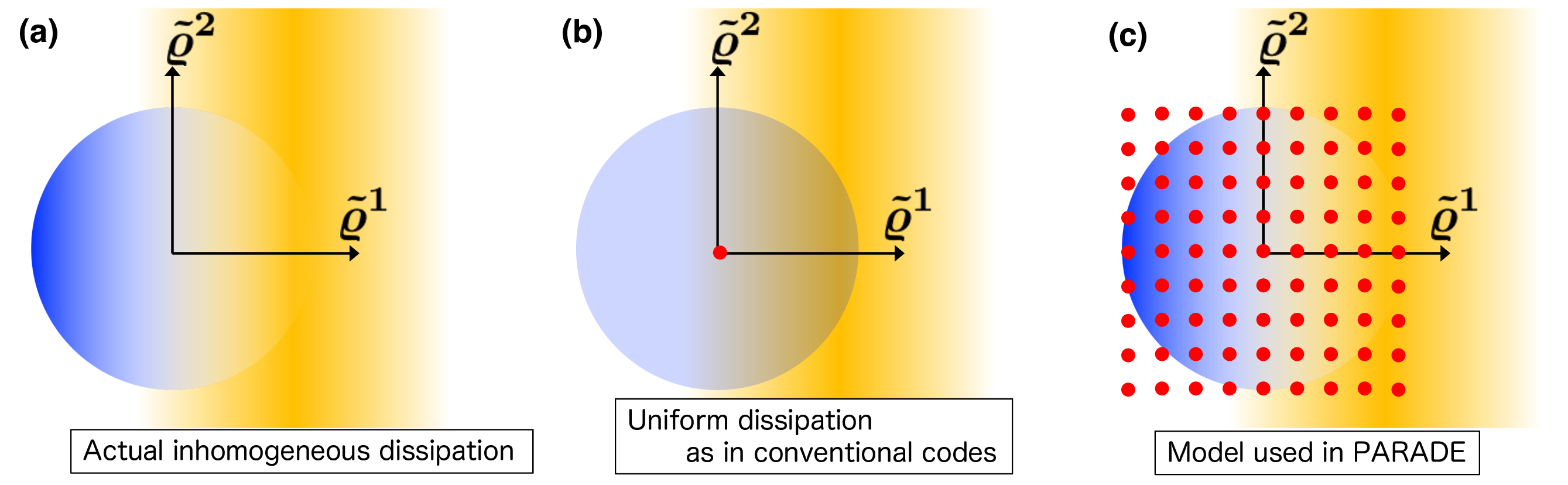}
\end{center}
\caption{Schematics of three models of resonant dissipation in an inhomogeneous medium. The wave beam propagates in the direction perpendicular to the figure. The intensity of the blue color denotes the intensity profile in the beam cross section. The intensity of the orange color denotes the local damping rate $\gamma$. The red points correspond to the locations at which $\gamma$ is actually calculated numerically. Figure (a) corresponds to the true inhomogeneous dissipation, in which case the field on the left dissipates more slowly than the field on the right (modulo diffraction). Figure (b) illustrates the homogeneous-dissipation model assumed in most quasioptical codes; in this model, the whole beam dissipates at the rate $\gamma$ evaluated at the beam center. Figure (c) illustrates the \parade dissipation model described in \Sec{sec:full}.
}
\label{fig:schematic}
\end{figure*}

\begin{figure*}
\begin{center}
\includegraphics[width=15.0cm,clip]{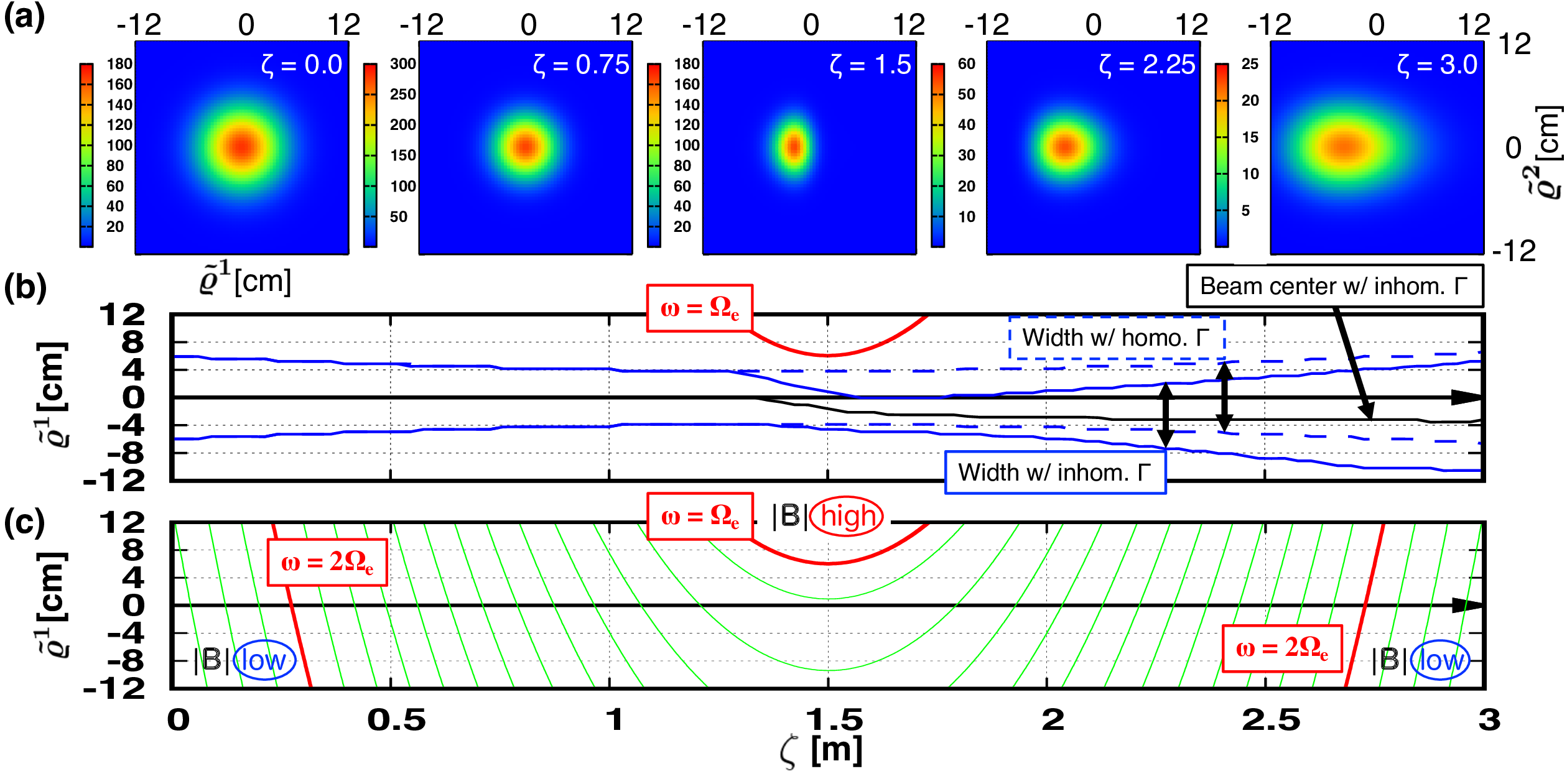}
\end{center}
\caption{Results of a test simulation of a wave beam passing near the electron cyclotron resonance within the model described in \Sec{sec:full}. Figure (a) shows the transverse cross sections of the beam intensity at $\zeta = 0.75$, $1.5$, $2.25$, and $3.0$~m. Figure (b) shows the evolution of the beam width (blue solid lines) and the trajectory of the beam center (black solid line). The beam width is determined numerically as the distance on which the amplitude $|a|$ drops one $e$-fold from its maximum along the $\tilde{\varrho}^\sigma$-axis. The beam center is defined as in \Eq{eq:moment-metod-1}. As a reference, the blue dashed lines show the corresponding width within the homogeneous-dissipation model [\Fig{fig:schematic}(b)]. Figure (c) shows isosurfaces of the magnetic-field strength [\Eq{eq:B}]. The locations of the first and second electron-cyclotron resonances are marked in red.}
\label{fig:partially}
\end{figure*}

In one scheme, we calculate $\vec{\Gamma}$ at each grid point using \Eq{eq:Gamma} as is [\Fig{fig:schematic}(c)]. Figure \ref{fig:partially} illustrates application of this model to a test simulation. Modeled there is the wave-beam propagation in hot-electron plasma with electron density $n = 1.0 \times 10^{19}$~m$^{-3}$, electron temperature $T = 2.0$~keV, and magnetic field $\{B_x, B_y, B_z\} = \{B \cos \theta, 0, B \sin \theta\}$, with $\theta = 80.0^\circ$,
\begin{gather}\label{eq:B}
B = B_0 \exp{
\bigg[- \bigg( \frac{x}{L_y}-1 \bigg)^2 - \bigg( \frac{x}{L_y}-1 \bigg)^2 \bigg]
},
\end{gather}
$B_0 = 7.3$~T, $L_x = 1.5$~m, and $L_y = 5.0$~m.
A single-mode O-wave beam is injected along the~$x$ axis from the origin and is initially assumed Gaussian, namely \cite{book:yariv},
\begin{multline}\label{eq:gb}
a^s = \sqrt{\frac{2}{\pi w_1 w_2}}\,\exp
\bigg[
- \frac{(\tilde{\varrho}^1)^2}{{w_1}^2} - \frac{(\tilde{\varrho}^2)^2}{{w_2}^2}
\\
+ \frac{i k (\tilde{\varrho}^1)^2}{2 R_1} + \frac{i k (\tilde{\varrho}^2)^2}{2 R_2}
+ \frac{i}{2} (g_1 + g_2)
\bigg],
\end{multline}
where
\begin{gather}
w_\mu \doteq w_{0,\mu} \sqrt{1 + \varsigma^{-2}},
\quad
R_\mu \doteq \mc{Z}_\mu (1 + \varsigma^2),
\\
g_\mu \doteq \tan^{-1} \varsigma,
\quad
\varsigma \doteq k w_{0,\mu}/(2\mc{Z}_\mu),
\end{gather}
and $k = 2\pi f/c$, with the focal lengths $\mc{Z}_1 = \mc{Z}_2 = 1.5$~m, the waist sizes $w_{0,1} = w_{0,2} = 4.0$~cm, and the wave frequency is $f = 77.0$~GHz, which corresponds to the vacuum wavelength $\lambda_0 \approx 4$~mm. The cold-plasma $\vec{\varepsilon}$ is used for $\vec{D}_H$, and the hot-plasma $\vec{\varepsilon}$ \cite{book:stix} is used for $\vec{D}_A$ with six cyclotron harmonics retained. In these settings, O--X conversion is insignificant, so the single-mode version of \parade \cite{ref:pop2} was used. Like all simulations reported in this paper, these simulations were done on a laptop with Intel Core\textsuperscript{TM}~i7-8569U processor.

Unlike within the homogeneous-dissipation model, the top side of the beam in \Fig{fig:partially}(b) is strongly distorted by inhomogeneous cyclotron damping. The beam narrows and experiences defocusing due to diffraction, and the location of the beam center
\begin{gather}\label{eq:moment-metod-1}
\tilde{\vec{\varrho}}_{\rm center} \doteq
\frac{\int |a^s|\, \tilde{\vec{\varrho}}\,\dd^2\tilde{\varrho}}{\int |a^s|\,\dd^2\tilde{\varrho}}
\end{gather}
shifts down relative to the horizontal line. Note that this shift (marked with black arrows) is unrelated to refraction and cannot be captured by most quasioptical codes. The ability of \parade to capture such shifts makes it particularly useful for modeling the propagation of wave beams grazing cyclotron resonances. Beams like that are typical in fusion experiments, which involve oblique injection from the mid-plane launcher or arbitrary injection from the top launcher.

\begin{figure}
\begin{center}
\includegraphics[width=8.2cm,clip]{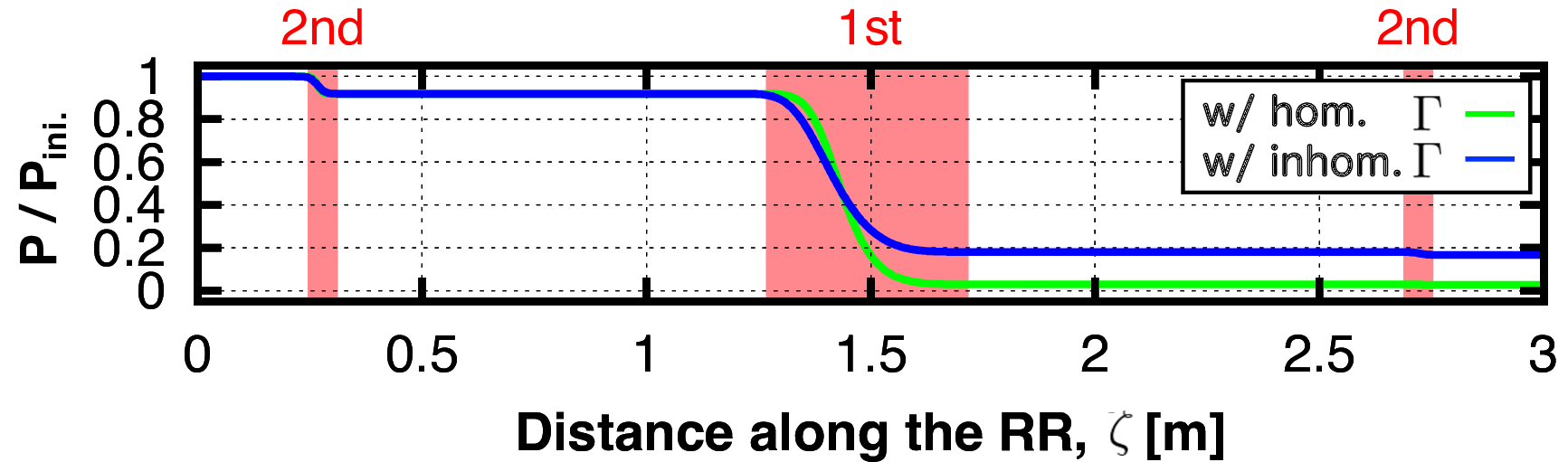}
\end{center}
\caption{The total power transported by the same beam as in \Fig{fig:partially}, in units of the input power. Green: homogeneous-dissipation model as in \Fig{fig:schematic}(b). Blue: inhomogeneous-dissipation model as in \Fig{fig:schematic}(c). The shaded regions denote areas of resonant dissipation at the first (figure center) and second (figure sides) cyclotron harmonics.}
\label{fig:remain}
\end{figure}

Figure \ref{fig:remain} shows the total beam power in the same simulation. The homogeneous-dissipation model predicts almost complete absorption of the beam power at the main cyclotron resonance. In contrast, the inhomogeneous-dissipation model predicts that the beam in fact retains about 20\% of its power. This is a significant difference, which is important in experiment \cite{note:nf1}.
%
%We will visit this topic in \cite{note:nf1}.

%--------------------------------------
\subsection{Approximate dissipation matrix}
\label{sec:apprx}

\begin{figure}
\begin{center}
\includegraphics[width=8.2cm,clip]{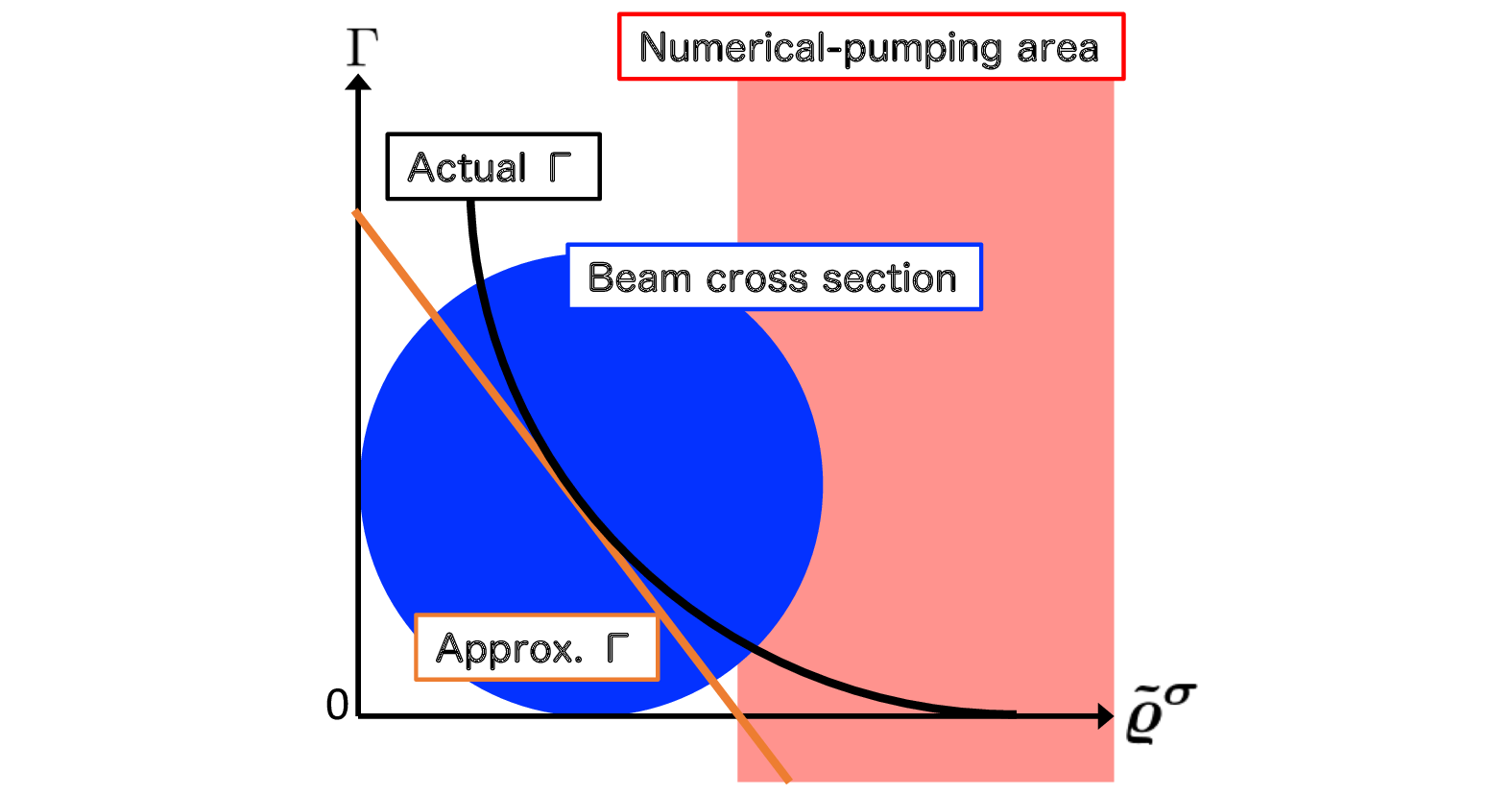}
\end{center}
\caption{A schematic of numerical pumping. The blue circle represents the beam cross section. The black curve and the orange line represent the actual dissipation-rate and the approximate dissipation-rate [as in \Eq{eq:GammaAppr}], respectively. Spurious amplification occurs in the red-shaded area, where the approximate dissipation-rate is negative. A similar discussion can also be found in \Ref{ref:balakin08a}.}
\label{fig:pump}
\end{figure}

Calculating $\vec{\Gamma}$ using \Eq{eq:Gamma} is computationally expensive and can be impractical, for example, for data analysis between discharges and for optimization of the launching geometry. Because of this, we also propose a simplified model as an alternative, which is as follows. Assuming that the beam width is smaller than the characteristic scale of $\vec{\Gamma}$, the latter can be Taylor-expanded to the first order in $\tilde{\vec{\varrho}}$. To do this, note that
\begin{gather}%\notag%\label{eq:Gaprx}
\vec{\Gamma}
= \vec{\Xi}_\star^+ \vec{D}_A(\vec{X} + \tilde{\vec{\varrho}}, \vec{K} + \tilde{\vec{\dk}}(\tilde{\vec{\varrho}})) \vec{\Xi}_\star,
\end{gather}
where $\tilde{\vec{\dk}}(\tilde{\vec{\varrho}}) \doteq \vec{k}(\vec{x}) - \vec{K}$ can be expressed as \cite{ref:pop2}
\begin{gather}\label{eq:dk}
\tilde{\dk}_\mu = - \frac{1}{V_\star^2}
\frac{\pd H_\star}{\pd \tilde{\varrho}^\sigma}
\frac{\pd H_\star}{\pd \tilde{\dk}_\mu}
\tilde{\varrho}^\sigma.
\end{gather}
This leads to
\begin{subequations}\label{eq:GammaAppr}
\begin{gather}
\vec{\Gamma} \approx \vec{\Gamma}_\star + \vec{\mc{G}}_{\star \sigma} \tilde{\varrho}^\sigma,\\
\vec{\mc{G}}_{\star \sigma}
= \frac{\pd \vec{\Gamma}_\star}{\pd \tilde{\varrho}^\sigma}
- \frac{1}{V_\star^2}
\frac{\pd H_\star}{\pd \tilde{\varrho}^\sigma}
\frac{\pd H_\star}{\pd \tilde{\dk}_\mu}
\frac{\pd \vec{\Gamma}_\star}{\pd \tilde{\dk}_\mu}.
\end{gather}
\end{subequations}
Because the matrix $\vec{\mc{G}}_{\star \sigma}$ has to be calculated only on the RR rather than at each grid point, this approach significantly speeds up calculations. We call it a first-order model. (Accordingly, the homogeneous-dissipation model used in other codes can be classified as the zeroth-model.) For single-mode beams, this approximation is equivalent to that used by Balakin \etal \cite{ref:balakin08a}, but our model extends to mode-converting beams as well.
When the first-order term exceeds the zeroth-order term, ``numerical pumping'' can occur (\Fig{fig:pump}; see also \Ref{ref:balakin08a}). To prevent this spurious effect in practical simulations, we introduce a cutoff:
\begin{gather}\label{eq:Gact}
\Gamma_{s's} \approx
\begin{cases}
    \Gamma_{\star s's} + \mc{G}_{\star s's \sigma} \tilde{\varrho}^\sigma,
    & \mc{G}_{\star s's \sigma} \tilde{\varrho}^\sigma \geq - \Gamma_{\star s's},
    \\
    0,
    & \mc{G}_{\star s's \sigma} \tilde{\varrho}^\sigma <    - \Gamma_{\star s's}.
\end{cases}
\end{gather}

\begin{figure*}
\begin{center}
\includegraphics[width=15.0cm,clip]{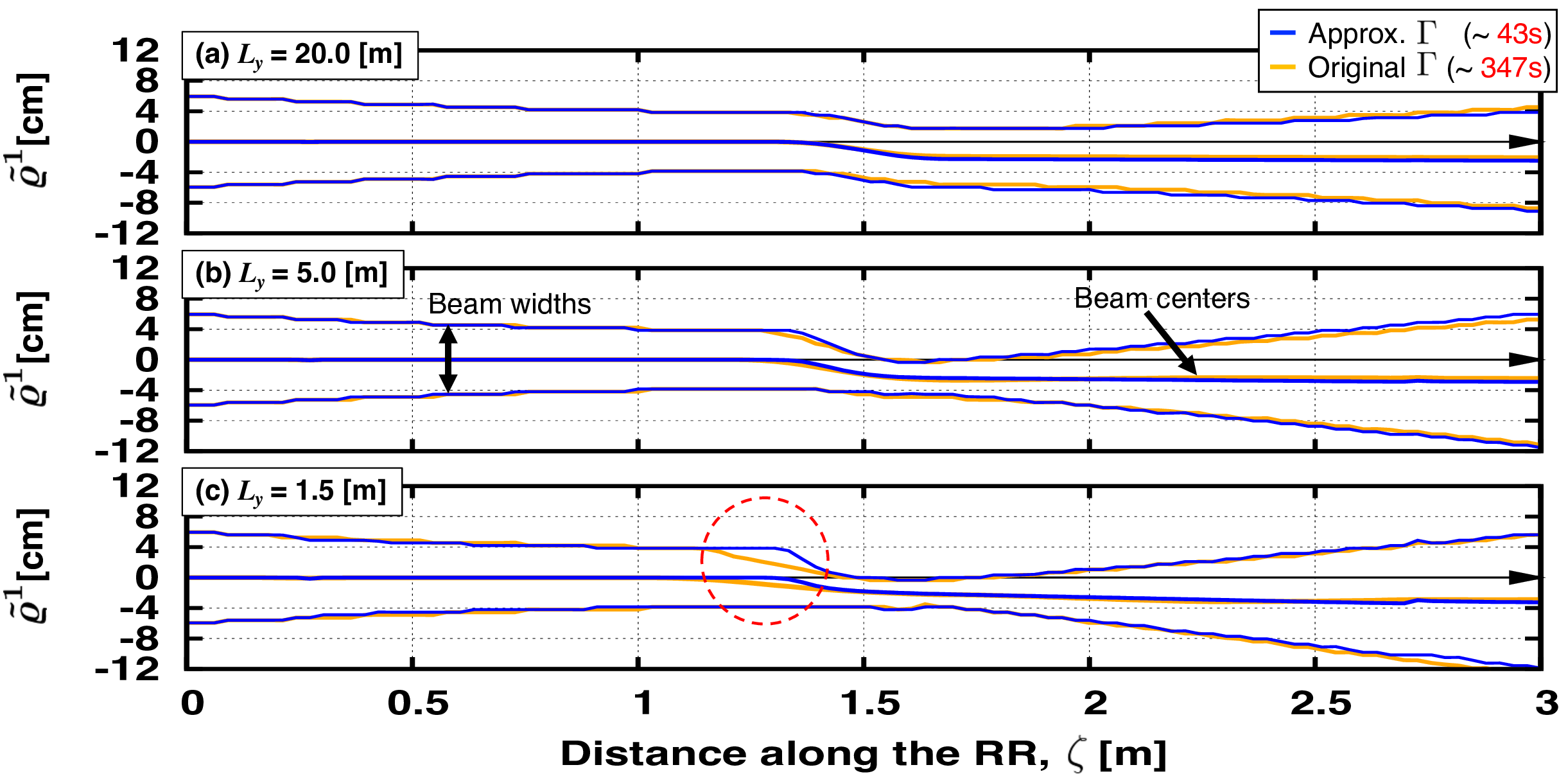}
\end{center}
\caption{Test-simulation results showing a comparison of the two models for $\vec{\Gamma}$ used in \parade. The curves illustrate the evolution of the beam widths and the beam center. Blue: approximate model \eq{eq:GammaAppr}. Orange: original model \eq{eq:Gamma}. The simulation setup is the same as in \Fig{fig:partially}, except three different values are used for the magnetic-field scale: (a)~$L_y = 20$~m, (b)~$L_y = 5$~m, and (c)~$L_y = 1.5$~m. The discrepancy between the two dissipation models is more pronounced at the smallest $L_y$, as marked with a red circle in figure~(c). The calculation using the approximate model is eight times faster.}
\label{fig:approx}
\end{figure*}

Figure \ref{fig:approx} demonstrates that the model \eq{eq:Gact} (blue curves) and the model \eq{eq:Gamma} (orange curves) are in reasonable agreement with test simulations. The simulation using an approximated $\Gamma$ is eight times faster than the one using the exact $\Gamma$ (43~s vs.\ 347~s), and in geometries of practical interest, the speed-up is anticipated to be even larger. However, the simplified model may not be sufficiently accurate when the magnetic-field scales are small enough, as seen in \Fig{fig:approx}(c).

%--------------------------------------
\subsection{Dissipation-driven mode conversion}
\label{sec:mc}

\begin{figure}
\begin{center}
\includegraphics[width=8.2cm,clip]{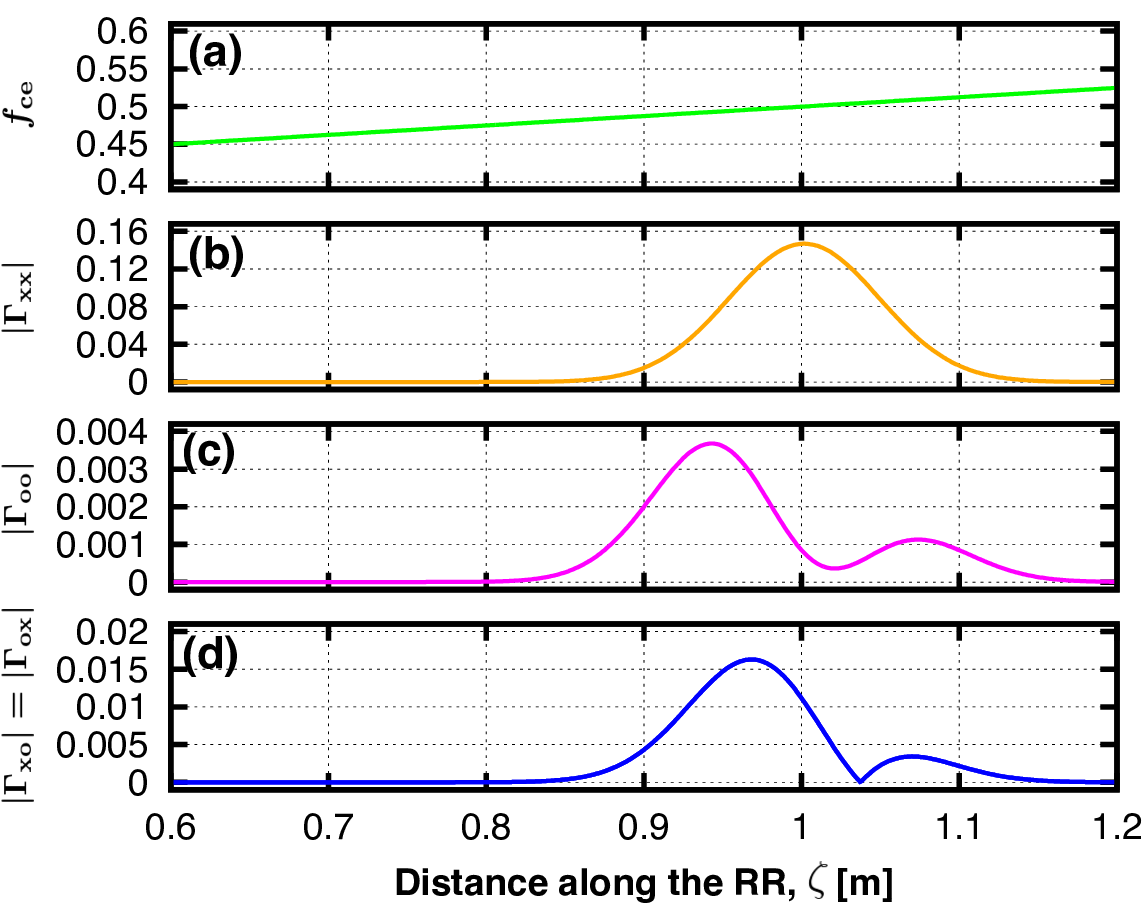}
\end{center}
\caption{Plasma parameters along the RR trajectory in a test simulation. Figure (a) shows the electron cyclotron frequency $f_{\rm ce}$ in units $f = 140.0$~GHz, which corresponds to the vacuum wavelength $\lambda_0 \approx 2$~mm. Figures (b)--(d) show the individual elements of $\vec{\Gamma}$. Here, the density is $n = 1.0 \times 10^{19}$~m$^{-3}$, the temperature is $T = 10$~keV, the magnetic field is $\{B_x, B_y, B_z\}=\{B\cos \theta, 0, B\sin \theta\}$, with $\theta = 85.0^\circ$ and $B = B_0 (x + x_0) / L_x$, where $B_0 = 2.5$~T, $x_0 = 3.0$~m, and $L_x = 4.0$~m. The wave is injected along the~$x$ axis from the origin.}
\label{fig:Gcomp}
\end{figure}

\begin{figure}
\begin{center}
\includegraphics[width=8.2cm,clip]{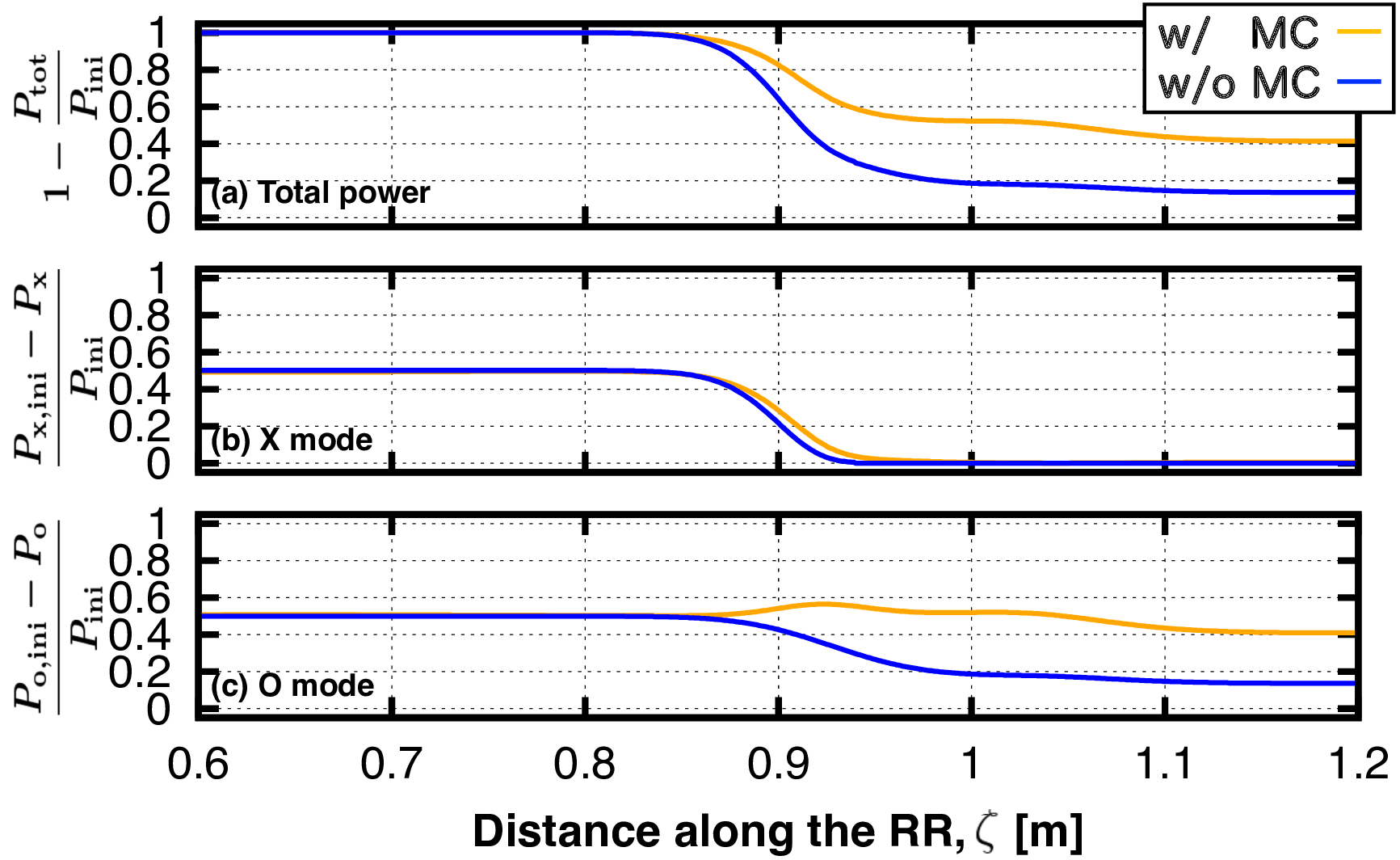}
\end{center}
\caption{\tored{The fractions of the remaining wave power in the same simulation as in \Fig{fig:Gcomp}. (a), (b), and (c) correspond to the result for total power: $1 - P_{\rm tot}/P_{\rm ini}$, X mode component: $(P_{\rm x,ini} - P_{\rm x})/P_{\rm ini}$, and O mode component: $(P_{\rm o,ini} - P_{\rm o})/P_{\rm ini}$, respectively. Here, $P_{\rm tot} \doteq P_{\rm x} + P_{\rm o}$ is the total absorbed power [\Eq{eq:PxPo}], and $P_{\rm ini} \doteq P_{\rm x,ini} + P_{\rm o,ini}$ is the total input power. The orange curves represent simulations where the dissipation-driven mode conversion is taken into account. For a reference, the blue curves represent simulations where mode conversion is ignored. A significant difference is appeared between both approach, especially for the O mode component $P_{\rm o}$, as seen from (c).}}
\label{fig:remainMC}
\end{figure}

The dissipation matrix $\vec{\Gamma}$ is generally nondiagonal (\Fig{fig:Gcomp}), so it couples different components of $\vec{\phi}$ in \Eq{eq:parab}, \ie causes mode conversion. This is particularly important when the dispersion relations of the two cold-plasma modes differ only slightly, so the coupling is strong. The corresponding applications include multi-pass heating at high harmonics, heating during the ramp-up phase, heating on medium- or small-size fusion devices, and also when off-axis heating is used to eliminate magnetic islands. In these cases, power absorption can be very different from that of a single-mode beam and thus cannot be properly modeled by codes that ignore mode conversion. In contrast, \parade is naturally suited to handle this problem.

To illustrate the effect of dissipation-driven mode conversion, we have performed a test simulation with the same parameters as in \Fig{fig:Gcomp}. The initial beam contains O~and X~modes in equal proportions and is Gaussian in shape [\Eq{eq:gb}], with $\mc{Z}_1 = \mc{Z}_2 = 2.0$~m and $w_{0,1} = w_{0,2} = 5.0$~cm. \tored{Figure \ref{fig:remainMC}~(a) shows the fraction of the remaining wave power, $1 - P_{\rm tot}/P_{\rm ini}$. Also, \Figs{fig:remainMC}~(b): $(P_{\rm x,ini} - P_{\rm x})/P_{\rm ini}$ and (c): $(P_{\rm o,ini} - P_{\rm o})/P_{\rm ini}$ show the remaining X- and O-components, respectively.} Here $P_{\rm ini}\tored{ = P_{\rm x,ini} + P_{\rm o,ini}}$ is the total input power, $P_{\rm tot} = P_{\rm x} + P_{\rm o}$ is total absorbed power, and $P_{\rm x}$ and $P_{\rm o}$ are defined as follows:
\begin{subequations}\label{eq:PxPo}
\begin{gather}
\left(
\begin{array}{c}
P_{\rm x}\\
P_{\rm o}
\end{array}
\right) \doteq \left(
\begin{array}{c}
P_{\rm xx} + P_{\rm xo}\\
P_{\rm ox} + P_{\rm oo}
\end{array}
\right),\\
P_{s's} = \int\,a^{s'} \Gamma_{s's} a^{s}\,\dd^2\tilde{\varrho}.
\end{gather}
\end{subequations}
\tored{The orange curves in each \Figs{fig:remainMC}~(a)-(c) represent} simulations where the dissipation-driven mode conversion is taken into account. For a reference, \tored{the blue curves represent} simulations where mode conversion (\ie the terms $P_{\rm xo}$ and $P_{\rm ox}$) is ignored.

The impact of the mode coupling on the \tored{total} power absorption is significant, as seen from the deviation of the orange curve from the blue curve in \Fig{fig:remainMC}\tored{~(a)}. \tored{Notably, as seen from \Fig{fig:remainMC}~(b) and (c), while X mode components with sufficiently high dissipation-rate are completely dissipated for both simulations, O mode component with mode conversion is increased unlike those without mode conversion.} Also note that although the polarization state of the wave was chosen here arbitrarily, calculating it for a practical experiment may also require \parade simulations, such as those described in \Ref{ref:pop3}.

%%%%%%%%%%%%%%%%%%%%%%%%%%%%%%%%%%%%%%%
\section{Conclusions}
\label{sec:conc}

Here, we report the first quasioptical simulations of wave beams in a hot plasma using the quasioptical code \parade (PAraxial RAy DEscription). This code \cite{ref:pop1, ref:pop2, ref:pop3} is unique in that it accounts for inhomogeneity of the dissipation rate across the beam and mode conversion simultaneously. We show that the dissipation-rate inhomogeneity shifts beams relative to their trajectories in cold plasma and that the two electromagnetic modes are coupled via this process, an effect that was ignored in the past. We also propose a simplified approach to accounting for the dissipation-rate inhomogeneity. This approach is computationally inexpensive and simplifies analysis of actual experiments. Our results lay the foundation for comparing \parade simulations with experimental data, as to be reported in our next paper \cite{note:nf1}.

\mbox{}

\section{Acknowledgments}
The work was supported by the U.S. DOE through Contract No.~DE-AC02–09CH11466. The work was also supported by JSPS KAKENHI Grant Number JP17H03514.

\bibliography{paper4}

\end{document}